# A NOTE ON TRANSFORMATIONS ON AUXILIARY VARIABLE IN SURVEY SAMPLING

Rajesh Singh and Mukesh Kumar

Department of Statistics, Banaras Hindu University,
Varanasi (U.P.), India
rsinghstat@yahoo.com

**ABSTRACT**

In this note, we address the doubts of Singh (2001) and Gupta and Shabbir (2008) on the transformations of auxiliary variables by adding unit free constants. The original contribution by Sisodia and Dwivedi (1981) is correct.

## 1. INTRODUCTION

The problem of estimating the population mean in the presence of an auxiliary variable has been widely discussed in finite population sampling literature. Ratio, product and difference methods of estimation are good examples in this context. In recent years, a number of research papers on ratio-type, exponential ratio-type and regression-type estimators have appeared, based on different types of transformations. Some important contributions in this area are due to Upadhyaya and Singh (1984), Pandey and Dubey (1988), Singh and Tailor (2003), Shabbir and Gupta (2005, 2006), Kadilar and Cingi (2004, 2006 a, b, c), Khosnevisan et al. (2007), Singh et. al. (2007), Gupta and Shabbir (2008), Singh et al. (2008) and Koyuncu and Kadilar (2008, 2009).

---

**Acknowledgements:** Thanks are due to Professor H.P. Singh, Vikram University, Ujjain, MP, India and Dr. Sarjinder Singh, Department of Mathematics, Texas A&M University-Kingsville, Kingsville, TX 78363, USA for a nice discussion on this topic.

Khoshnevisan *et al.* (2007) suggested a family of estimators for estimating population mean $\bar{Y}$ as

$$t = \bar{y}\left[\frac{a\bar{X}+b}{\alpha(a\bar{x}+b)+(1-\alpha)(a\bar{X}+b)}\right]^g \qquad (1.1)$$

where $a(\neq 0)$, $b$ are either real numbers or the functions of the known parameters of the auxiliary variable x such as standard deviation ($\sigma_x$), coefficient of variation ($C_x$), skewness ($\beta_1(x)$), kurtosis ($\beta_2(x)$) and correlation coefficient ($\rho$).

Singh (2001) and Gupta and Shabbir (2008) raised doubt about the utility of some of these transformations. Singh (2001) pointed out that coefficient of variation and coefficient of skewness was used by different authors in additive form to the sample and population means of the same character. He further observed that that coefficient of variation and coefficient of skewness are unit free constant, their additions may not be justified.

## 2. CLARIFICATION

Here we address the doubts of Singh (2001) and Gupta and Shabbir (2008). Suppose $x_1, x_2, ..., x_n$ are the weights in kilogram (kg.) of $n$ persons. Following Sisodia and Dwivedi (1981), we transform the auxiliary variable in the population and the sample as follows:

$$\frac{\bar{X}+C_X}{\bar{x}+C_X} \qquad (2.1)$$

Note that in the expression (2.1), we have

$$\bar{X}(\text{kg}) + 1(\text{kg})C_x \qquad (2.2)$$

For example, in particular, if the population mean and the population coefficient of variation are 5 kg and 0.3, respectively.

This means, we are writing

$$\bar{X}(\text{kg}) + 1(\text{kg})C_x = 5(\text{kg}) + 1(\text{kg})(0.3) = 5.3(\text{kg}) \qquad (2.3)$$

Note here that $C_x$ multiplied by one has the same unit of measurement as the population mean $\bar{X}$, so there is no problem in using coefficient of variation, and hence any other unit free measure such as coefficient of skewness in an additive form.

In the same way, one could think a transformation:

$$\frac{\sigma_x^2 + \alpha}{s_x^2 + \alpha} \quad (2.4)$$

Again note that in the expression (2.4), we have

$$\sigma_y^2 \left(\text{kg}^2\right) + (1\text{kg} \times 1\text{kg})\alpha \quad (2.5)$$

For example, in particular, if the population variance and the constant $\alpha$ are 25 kg$^2$ and 0.3, respectively.

This means, we are writing

$$25\left(\text{kg}^2\right) + (1\text{kg} \times 1\text{kg})(0.3) = 25.3 kg^2 \quad (2.6)$$

These kinds of adjustments of units of measurements are a common practice in elementary mathematics. For example, let

$$f(x) = \begin{cases} 1 & if \quad a < x < b \\ 0 & otherwise \end{cases} \quad (2.7)$$

The area under the function will be $(b - a)$ which is in fact the distance between two points $a$ and $b$ on a real line, say $(b - a)\text{cm}$. Thus, to make $(b - a)\text{cm}$ as area, we have to multiply by 1 unit of measurement of length ( say, 1cm). The required area will be: $(b - a)\text{cm} \times 1\text{cm} = (b - a)\text{cm}^2$

Now take another counter example, if we integrate speed (cm/sec) of an object with respect to time (sec), then we get distance in cm. Now distance is not area, as expected after integrating a function, but it is distance in cm. One could also think that $\pi$ is a constant and irrational number, but has radians as units of measurements.

We conclude that any such unit free constant used in transformation of auxiliary variable inherits the character of adopting the units of the type of transformation being applied on the auxiliary variable.